\documentclass[letterpaper,11pt]{article}
\usepackage[latin1]{inputenc}
\usepackage[spanish,english,activeacute]{babel}%
\usepackage[letterpaper, left=2cm, top=2cm, right=1.5cm, bottom=1.5cm]{geometry}
\usepackage{amsmath,amsthm,amsfonts,amssymb,bm}
\usepackage{setspace}
\usepackage[dvips,dvipdfm]{graphicx}
\usepackage{graphicx,psfrag,float,here}
\usepackage{dcolumn}
\usepackage{caption2}
\usepackage[superscript]{cite}
\usepackage{pst-all}
\psset{xunit=1cm,unit=1cm,fillcolor=white,linewidth=.3mm,arrowsize=5pt,coilarm=.25cm,coilwidth=.25cm}
\newcommand{\ket}[1]{\ensuremath{|#1\rangle}}
\renewcommand{\Im}{\ensuremath{\textnormal{Im}}}
\newcommand{\bra}[1]{\ensuremath{\langle#1|}}
\DeclareRobustCommand\openone{\leavevmode\hbox{\small1\normalsize\kern-.33em1}}
\newcommand{\tr}[2][\mbox{}]{\textrm{tr}_{#1}\left[#2\right]}
\newcommand{\braket}[2]{\langle#1|#2\rangle}
\title{\bf\MakeUppercase{Discordia Cuántica de Osciladores $f-$deformados Bipartitos Entrelazados}}
\author{E. Castro,\thanks{Envío de correspondencia: ecastro@usb.ve} \and S. D\'{i}az-Sol\'orzano, \and R. G\'{o}mez, \and A. Zambrano \and C. L. Ladera}
\date{\begin{minipage}{.8\textwidth}\centering\small%
Grupo de Información y Comunicación Cuántica, Departamento de Física, Universidad Sim\'on
Bol\'ivar, Sartenejas, Edo. Miranda 89000, Venezuela.
\end{minipage}}

\begin{document}
\maketitle%
\renewcommand{\abstractname}{Resumen}
\begin{abstract}\noindent%
Las correlaciones cuánticas en sistemas compuestos son de gran importancia, y son un recurso
fundamental para el desarrollo de protocolos en computación e información cuántica. En este
trabajo, utilizando el formalismo de los osciladores f-deformados, se construyen estados
coherentes puros no lineales usando la función de deformación de Barut-Girardello. También se
construye la extensión a estados mixtos utilizando los estados tipo Werner. Para los estados
construidos, se estudia la dependencia del grado de entrelazamiento y de la discordia cuántica, en
función de las deformaciones de los estados, y su grado de mezcla.
\end{abstract}

\section*{Introducción}
\mbox{}\\*[-1cm]
\par\noindent
 El entrelazamiento es una de las correlaciones que no tiene contraparte clásica, y su
efecto marca una gran diferencia en los comportamientos clásicos y cuánticos de sistemas
compuestos. Schrödinger\cite{schrodinger1935-1936} fue el primero en darse cuenta de este efecto,
al percatarse que el conocimiento de la información maximal del estado de un sistema compuesto por
dos particiones no garantiza el total conocimiento de la información arrojado por los estados de
cada partición que cons\-tituye al sistema. El entrelazamiento ha sido usado como recurso en
procesos de teleportación\cite{Bennett1993}, codificación superdensa\cite{Bennett1992}, clonación
cu\'{a}ntica\cite{Gisin1997}, algoritmos cu\'{a}nticos\cite{Shor1994,Grover1997}, criptografía
cu\'{a}ntica\cite{Vidal2003}, y tecnologías de computación cu\'{a}ntica\cite{Bennett1996}, entre
otros. Sin embargo, el entrelazamiento cu\'{a}ntico no es la \'{u}nica correlación cu\'{a}ntica
\'{u}til para el procesado cuántico de información (PCI). La Discordia Cu\'{a}ntica (DC)
introducida inicialmente por Ollivier y Zurek\cite{Ollivier2001}, y al mismo tiempo y de forma
independiente por Henderson y Vedral\cite{Henderson2001}, mide otro tipo de correlación que la
planteada por el entrelazamiento, que incluso puede ser distinta de cero a\'{u}n en aquellos casos
en los cuales no existe entrelazamiento, lo cual ocurre para una clase muy particular pero amplia
de estados separables\cite{ferraro2010}. La DC se define como la diferencia entre la información
mutua del sistema y la información mutua obtenida despu\'{e}s de medir sobre una de las
particiones de dicho sistema. Es decir, al medir localmente sobre las distintas partes del
sistema, no se obtiene la información del estado original antes de la medida, ya que parte de ésta
es inaccesible locamente; razón por la cual la DC se puede interpretar como una medida de la
fracción de información mutua a pares que es localmente
inaccesible,\cite{Ollivier2001,Henderson2001} o como la diferencia entre la información contenida
en todo el sistema y la que se obtiene al perturbar midiendo localmente uno de sus constituyentes
\cite{Ren2012}.
\\
\par\noindent  El entrelazamiento en variables continuas se ha presentado como una t\'{e}cnica
prometedora para codificar y manipular la información cu\'{a}ntica.\cite{Cerf} En particular, los
estados coherentes entrelazados\cite{Sanders2012}, a dos modos, presentan oscilaciones y
estadística fotónica más robustas que para los estados bi-fotónicos entrelazados
polarizados\cite{Bukhari2011yZheng2007yAshfag2006}. Recientemente, Hosseimi y
co\-la\-bo\-ra\-do\-res\cite{Hosseimi2014}, han usado la técnica de detección homodina para
encontrar experimentalmente la DC en sistemas de variable continua; usando estados Gaussianos y no
Gaussianos obtienen DC distintos de cero a\'{u}n para un valores pequeños de los par\'{a}metros
involucrados. De aquí la necesidad de cuantificar y comparar correlaciones cuánticas
(entrelazamiento y DC) para estados con variable continua.
\\
\par\noindent En el presente trabajo, enfocaremos el estudio de la DC y el entrelazamiento de formación
(EoF) para estados con variable continua, tomando los estados coherentes bipartitos no ortogonales
$f-$deformados tipo Bell, y su mezcla con el estado máximamente mixto, tipo cuasi-Werner. Una
parte de éste estudio fue presentado en una conferencia previa\cite{Castro2013}, por lo que el
presente trabajo muestra algunos nuevos resultados del EoF y la DC en forma analítica, para los
estados antes mencionados, introduciendo mejoras, co\-rrec\-cio\-nes y optimizaciones con respecto
a los publicados en dicha conferencia. En el presente trabajo se determinó una expresión analítica
de la DC, la cual ya está maximizada en su entropía condicional, y se determinó el EoF en forma
analítica. Esta publicación est\'{a} estructurada en tres partes. En la primera, se hace una
introducción a los estados coherentes no li\-nea\-les $f$-deformados. En la segunda parte, se
presenta una descripción de la DC, y se muestran resultados analíticos de dicha medida de
correlación; así como de la concurrencia y el EoF. En la tercera parte, se discuten y analizan los
resultados obtenidos.

\section*{Osciladores $f-$deformados}
\mbox{}\\*[-1cm]
\par\noindent%
Los estados coherentes fueron descubiertos inicialmente por Scr\"{o}dinger\cite{schrodinger1926}
cuando determinó la función de onda dependiente del tiempo para un oscilador armónico; ya que
éstos deben estar en completo acuerdo con el movimiento oscilatorio clásico. Posteriormente, esto
estados fueron introducidos en la óptica cuántica por Glauber\cite{Glauber1963}, mediante el
estudio de la coherencia de los campos de radiación cuantizados, él demostr\'{o} que los estados
coherentes
$|\alpha\rangle$ pueden ser obtenidos a partir de cualesquiera de las si\-guien\-tes premisas: %
(I) Son los autovectores del operador de aniquilación bosónico $\hat{a}$, cuyo autovalor $\alpha$
es cualquier número complejo, comúnmente llamado variable continua. Es decir,
$\hat{a}|\alpha\rangle=\alpha|\alpha\rangle$. %
(II) Se obtienen de a aplicar el operador desplazamiento
$D\left(\alpha\right)=\exp\left(\alpha\hat{a}^\dag-\alpha^*\hat{a}\right)$ sobre el vacío
cu\'{a}ntico \ket{0} del oscilador arm\'{o}nico, es decir, $|\alpha\rangle = D(\alpha) \ | 0
\rangle $, siendo $ {\hat{a}}^{\dag}$  el operador de creaci\'{o}n bosónico. %
(III) Son estados cu\'{a}nticos para los cuales el producto de las dispersiones del operador
momento y posición presentan el mínimo valor de incertidumbre, es decir, en dichos estados se
cumple que $\left( \triangle q \right)_{\alpha}^{2} \left(\Delta p \right)_{\alpha}^{2} = \hbar /
4$, siendo $q$ y $p$ los operadores posición y momentum, respectivamente.
\\
\par\noindent Man'ko y colaboradores\cite{Manko1997} introdujeron los osciladores $f-$deformados,
como una generalizaci\'{o}n de los $q-$osciladores. Los $f-$osciladores describen oscilaciones con
un tipo específico de no linealidad; para los cuales la frecuencia de oscilación depende de la
e\-ner\-gía del oscilador. Estos osciladores no lineales, a un solo modo, se obtienen a partir de
la deformaci\'{o}n de los operadores de aniquilaci\'{o}n y creaci\'{o}n bos\'{o}nicos, de acuerdo
a las siguientes relaciones algebraicas
\begin{subequations}\label{eq:1}
\begin{align}
\label{eq:1a}%
\hat{A}        &\equiv \hat{a}f(\hat{n})= f(\hat{n}+1)\hat{a}\,,\\
\label{eq:1b}%
\hat{A}^{\dag} &\equiv f(\hat{n}){\hat{a}}^{\dag}=\hat{a}^{\dag}f(\hat{n}+1)\,,
\end{align}
\end{subequations}
donde $ \hat{n} = {\hat{a}}^{\dag} \hat{a}$, es el operador de n\'{u}mero bos\'{o}nico, y su
acci\'{o}n sobre las bases del espacio de Fox es
\begin{subequations}\label{eq:2}
\begin{align}
\label{eq:2a}%
\hat{A}\ket{n}&=\sqrt{n}f(\hat{n})\ket{n-1},\\
\label{eq:2b}%
\hat{A}^{\dag}\ket{n}&=\sqrt{n+1}f(\hat{n}+1)\ket{n+1}.
\end{align}
\end{subequations}
Los operadores bos\'{o}nicos deformados, $\hat{A}$ y ${\hat{A}}^{\dag}$, difieren de los
operadores bos\'{o}nicos usuales, $\hat{a}$ y ${\hat{a}}^{\dag}$, por una funci\'{o}n de
deformaci\'{o}n $f(\hat{n})$, siendo ésta una función continua, real y positiva tal que $f(0)=1$.
Las relaciones de conmutaci\'{o}n entre los operadores deformados est\'{a}n dadas por
\begin{equation}\label{eq:3}
[\hat{n},\hat{A}]=-\hat{A},             \hspace{.2cm}%
[\hat{n},\hat{A}^\dag]=-\hat{A}^{\dag}, \hspace{.2cm}\textrm{y}\hspace{.2cm}%
[\hat{A}, {\hat{A}}^{\dag}]=1+\phi(\hat{n}),
\end{equation}
donde $\phi(\hat{n})= (\hat{n}+1) \ f^{2}(\hat{n}+1) - \hat{n} \ f^{2}(\hat{n}) - 1$. La
deformaci\'{o}n es fija cuando se define la funci\'{o}n $f(\hat{n})$, y los estados coherentes se
recuperan cuando $f(\hat{n}) = 1$. Man'ko y colaboradores\cite{Manko1997}  introdujeron los
estados coherentes $f-$deformados ${|\alpha\rangle}_{A}$, de manera análoga a los estados
coherentes; es decir, como los autoestados del operador de aniquilaci\'{o}n deformado $\hat{A}$.
Por lo que, de la ecuación de autovalores $\hat{A} {|\alpha\rangle}_{A} = \alpha
{|\alpha\rangle}_{A}$ conduce a:
\begin{subequations}\label{eq:4}
\begin{gather}
\label{eq.4a}%
\ket{\alpha}_{A}=N_{A}\sum_{n=0}^{\infty} \frac{\alpha^{n}}{\sqrt{n!}f(n)!}\,\ket{n},\\
\label{eq:4b}%
N_{A}=\left[\sum_{n=0}^{\infty}\frac{|\alpha|^{2n}}{ \ n! \ f^{2}(n)!}\right]^{- 1/2},
\end{gather}
\end{subequations}
donde $N_{A}$ es la constante de normalizaci\'{o}n, $|\alpha|^2$ es el número medio de fotones, y
$ f(n)! = f(0) f(1) f(2)\cdots f(n)$. Estos estados poseen propiedades no cl\'{a}sicas como
compresi\'{o}n y antibunching.\cite{Matos1996} Debido a que el conmutador de $\hat{A}$ con
${\hat{A}}^{\dag}$ no es un n\'{u}mero (véase la Ec.~\ref{eq:3}) el operador des\-plazamiento
$\hat{D}(\alpha)$ no puede ser escrito como el producto de exponenciales, ni tampoco es unitario.
Recamier y colaboradores\cite{Matos1996} demuestran que los estados coherente $f-$deformado
${|\alpha\rangle}_{D}$ pueden ser obtenidos a partir de un operador desplazamiento $f-$deformado
$\hat{D}(\alpha,f)$, cuando se hace actuar sobre el estado vacío \ket{0} del espacio de Fox; en
otras palabras, ${|\alpha\rangle}_{D} = \hat{D}(\alpha,f) | 0 \rangle $. El operador
desplazamiento deformado puede ser escrito de la forma
\begin{equation}\label{eq:5}
{\hat{D}(\alpha,f)=\exp\left[\alpha {\hat{A}}^{\dag} \right]\exp\left[\alpha^{*}
\hat{A}\right]\exp\left[-\tfrac{|\alpha|^{2}}{2} \left( 1 + \phi(\hat{n}) \right) \right]. }
\end{equation}
Este operador desplazamiento es aproximadamente unitario, y desplaza tanto al operador $\hat{A}$
como al o\-pe\-ra\-dor ${\hat{A}}^{\dag}$ si y solo si $|\alpha|^{2}\phi(\hat{n})/2\ll1$. Este
operador desplazamiento satisface las relaciones
\begin{subequations}\label{eq:6}
\begin{gather}
\label{eq:6a}%
\hat{D}(-\alpha,f)={\hat{D}}^{-1}(\alpha,f)=\hat{D}^{\dag}(\alpha,f),
\\
\label{eq:6b}%
{\hat{D}}^{\dag}(\alpha,f)\hat{A}\hat{D}(\alpha,f)=\hat{A}+\alpha\left(1-\phi(\hat{n})\right),
\\
\label{eq:6c}%
{\hat{D}}^{\dag}(\alpha,f)\hat{A}^{\dag}\hat{D}(\alpha,f)=\hat{A}^{\dag}+\alpha^{*}\left(1-\phi(\hat{n})\right).
\end{gather}
\end{subequations}
También obedece la relaci\'{o}n de semigrupo, esto es,
\begin{equation}\label{eq:7}
{\hat{D}(\alpha,f)\hat{D}(\beta,f)=
\hat{D}(\alpha+\beta,f)\exp\left[i\Im\left(\alpha{\beta}^{*}\right)\left(1+\phi(\hat{n})\right)\right].
}
\end{equation}
Los estados coherentes $f-$deformados normalizados, obtenidos a partir del operador desplazamiento
$\hat{D}(\alpha,f)$ tienen la forma
\begin{subequations}\label{eq:8}
\begin{align}
\label{eq:8a} {|\alpha\rangle}_{D}
&=\hat{D}(\alpha,f)\ket{0}=N_{D}\sum_{n=0}^{\infty}\frac{\alpha^{n}f(n)!}{\sqrt{n}!}\ket{n},\\
\label{eq:8b} N_{D} &=\left[\sum_{n=0}^{\infty}\frac{|\alpha|^{2n}f^{2}(n)!}{n!}\right]^{- 1/2}.
\end{align}
\end{subequations}
Los dos tipos de estados deformados ${|\alpha\rangle}_{A}$ y ${|\alpha\rangle}_{D}$ presentan una
evoluci\'{o}n cu\'{a}ntica similar, pero sus estadística son diferentes\cite{Recamier2008}.
\\
\par\indent
Los estados coherentes asociados con el álgebra de Lie $SU(1,1)$ fueron construidos por Barut y
Girardello \cite{Barut} y por Gilmore y Perelomov\cite{Perelomov,Honarasa}. Las funciones no
lineales de deformaci\'{o}n asociadas con estos estados $SU(1,1)$ son $f_{BG}(n,\kappa)=\sqrt{n +
2\kappa - 1}$ y $f_{GP}(n,\kappa)=\sqrt{n+2\kappa-1}$, respectivamente, con $\kappa=1/2, 1,
3/2,\cdots$. Con la finalidad de ilustrar la influencia de la deformación la influencia de la
deformación sobre la DC y el EoF, en el presente trabajo utilizaremos solamente la función
$f_{BG}$, como nuestra función no lineal de deformaci\'{o}n, de los estados $f-$deformados
${|\alpha\rangle}_{D}$.
\\
\par\noindent En el presente trabajo consideraremos a los estados coherentes $f-$deformados como un recurso para
codificar informaci\'{o}n o desarrollar protocolos cuánticos. Para ello consideraremos las bases
ortogonales
\begin{equation}\label{eq:9}
|\pm\rangle = N_{\pm} \left( {|\alpha\rangle}_{D} \pm {|-\alpha\rangle}_{D} \right) \ ,
\end{equation}
conocidas como las bases ortonormales deformadas, impares y pares, tipo Gato de Schr\"{o}dinger,
respectivamente, donde $N_{\pm}=\left(2\pm2\;\mbox{}_D\braket{\alpha}{-\alpha}_D\right)^{-1/2}$
son las constantes de normalizaci\'{o}n. Consideraremos adem\'{a}s el estado puro deformado, no
ortogonal, tipo Bell de la forma
\begin{equation}\label{eq:10}
|{\psi}^{+}\rangle = n_{+} \left( {|\alpha\rangle}_{D} \otimes {|\alpha\rangle}_{D} \pm
{|-\alpha\rangle}_{D} \otimes {|-\alpha\rangle}_{D} \right),
\end{equation}
donde la constante de normalizaci\'{o}n viene dada por $n_{+}=\left(2+2\;|\mbox{}_D
\langle\alpha|-\alpha\rangle_{D}|^{2}\right)^{-1/2}$. Considerando las bases ortonormales
$|\pm\rangle$, de forma tal que los estados $|{\psi}^{+}\rangle$ se pueden escribir de la forma
\begin{equation}\label{eq:11}
|{\psi}^{+}\rangle = \frac{n_{+}}{2} \left[ \frac{|+,+\rangle}{\left( N_{+}\right)^{2}} +
\frac{|-,-\rangle}{\left( N_{-}\right)^{2}}\right],
\end{equation}
Los estados cuasi-Werner $\rho_{W}$ son construido a partir de la combinación lineal entre el
operador identidad $\openone_{4}$, de orden 4, y el proyector construido con el estado dado en
\eqref{eq:11}. La interpretación física de este estado corresponde a la mezcla estadística entre
los estados $\tfrac{1}{4}\openone_{4}$ y $\ket{\psi^+}\bra{\psi^+}$ con probabilidad $(1-p)$ y
$p$, respectivamente. Esto es,
\begin{equation}\label{eq:12}
\rho_{W}(\alpha,f(n),p)=\frac{(1-p)}{4}\openone_{4} + p |{\psi}^{+}\rangle\langle{\psi}^{+}| \ .
\end{equation}
\section*{Entrelazamiento de formación y Discordia cuántica}
Una buena medida para cuantificar el entrelazamiento de un estado puro $\psi$ es la entropía de
von-Neumann (definida como $S(\rho)=-\rho\log_2(\rho)$, en unidades de bits), la razón se debe a
la posibilidad de convertir un estado puro entrelazado en otro a partir de una cierta cantidad de
estados singletes\cite{Wooters}, dicha cantidad es proporcional a la entropía del estado asociado
a una de sus particiones. Así, que el entrelazamiento de formación para un estado puro viene dado
por
\begin{equation}\label{eq:20}
EoF(\psi)=S(\rho_A)=S(\rho_B)
\end{equation}
Siendo $\rho_A=\tr[B]{\ket{\psi}\bra{\psi}}$ y $\rho_B=\tr[A]{\ket{\psi}\bra{\psi}}$ los estados
reducidos. Es claro que el entrelazamiento de formación vista como la entropía para un estado puro
no cambia antes o\-pe\-ra\-cio\-nes locales, por lo que no es posible la creación o destrucción de
entrelazamiento usando transformaciones unitarias locales. No obs\-tan\-te, la entropía de
von-Neumann no es una buena medida del grado de entrelazamiento para estados mixtos, la razón es
que existen estados no entrelazados cuya entropía de algunas de sus partes puede no anularse. Para
cuantificar el grado de entrelazamiento de un estado mixto Wooters\cite{Wooters} propone al
promedio del entrelazamiento de las particiones que componen al sistema y luego minimizando sobre
toda la composici\'{o}n, es decir:
\begin{equation}\label{eq:19}
EoF(\rho)= \min \sum_{i} P_{i} E\left(\psi_{i}\right) ,
\end{equation}
siendo $E\left(\psi\right)=-\tr{\rho_{A}\log_{2}\rho_{A}}=-\left[ \rho_{B} \log_{2} \rho_{B}
\right]$ es la entrop\'{i}a asociada a cada una de las partes del sistema bipartito. En cualquier
caso (puro o mixto) el EoF se puede escribir como
\begin{equation}\label{eq:21}
Eo{F}(\rho)= h \left(\frac{1+\sqrt{1-C^{2}}}{2}\right) \ ,
\end{equation}
\noindent siendo $h(x) = - x  \log_{2}  x - (1-x)  \log_{2}  (1-x)$ la funci\'{o}n de entrop\'{i}a
binaria. El par\'{a}metro real y positivo $C$ recibe el nombre de \emph{concurrencia}, y viene
definida para estados puros y mixtos como $C=|\braket{\psi}{\tilde\psi}|$ y
$C=\max\left\{0,\sqrt{\lambda_{1}}-\sqrt{\lambda_{2}}-\sqrt{\lambda_{3}}-\sqrt{\lambda_{4}}
\right\}$, respectivamente. Siendo los $\lambda_{i}$ los autovalores, en orden decreciente, de la
matriz positiva y hermítica $R=\sqrt{\sqrt{\rho} \ \tilde{\rho} \ \sqrt{\rho}}$, o
alternativamente, son las raíces de los valores propios de la matriz no hermítica $\rho
\tilde{\rho}$, y $\ket{\tilde\psi}$ así como $\tilde{\rho}$ resultan de aplicar la operaci\'{o}n
"spin-flip" sobre el estado puro $\ket{\psi}$ y el estado mezcla $\rho$, es decir,
$\ket{\tilde{\psi}}=\sigma_{y}\otimes\sigma_{y}\ket{\psi^\star}$ y $\tilde{\rho}=
\left(\sigma_{y}\otimes\sigma_{y}\right){\rho}^{\star}\left(\sigma_{y}\otimes\sigma_{y}\right)$,
respectivamente. Siendo \ket{\psi^\star} y ${\rho}^{\star}$ el conjugado de los estados puro y
mixto \ket{\psi} y $\rho$, respectivamente. Para los estados tipo-Werner indicados en
\eqref{eq:12} se tiene que la concurrencia es
\begin{equation}\label{eq:22}
C=\begin{cases}%
\sqrt{\lambda_+}-2\sqrt{\lambda_0}-\sqrt{\lambda_-}
&\textrm{si\space}q<\tfrac{1}{4}\\
\sqrt{\lambda_0}-2\sqrt{\lambda_+}
&\textrm{si\space}q=\tfrac{1}{4}\\
\sqrt{\lambda_-}-2\sqrt{\lambda_0}-\sqrt{\lambda_+}
&\textrm{si\space}q>\tfrac{1}{4}\\
\end{cases}
\end{equation}
Donde
\begin{subequations}\label{eq:23}
\begin{align}
q&=\frac{1+3p}{4}
\\
\label{eq:23a}%
\lambda_0&=\left(\frac{1-q}{3}\right)^2
\\
\lambda_\pm&={\scriptstyle \tfrac{q(1-q)}{3}+\tfrac{1-4q}{18}C_\psi\left[C_\psi\pm
\sqrt{(1-4q)^2C_\psi+12q(1-q)}\right]}
\end{align}
\end{subequations}
Siendo $C_\psi=n_+^2/2N_+^2N_-^2$ la concurrencia del estado puro \eqref{eq:11}.
\\
\par\indent
Para el cálculo de la DC la informaci\'{o}n mutua cu\'{a}ntica del sistema $AB$ viene dada
por\cite{Ollivier2001,Henderson2001}
\begin{equation}\label{eq:13}
I\left(\rho_{AB}\right)=S\left(\rho_{A}\right) + S\left(\rho_{B}\right) - S\left(\rho_{AB}\right).
\end{equation}
Mientras que la información mutua después de un conjunto de medidas proyectivas locales en la
partición $B$, tipo von Neumann, esto es, $\{\Pi_{B}^{j}\}=\{\openone_2\otimes
|{j}_{B}\rangle\langle{j}_{B}|\}$. El estado final de todo el sistema, una vez realizada la medida
sobre $B$, ser\'{a}
\begin{equation}\label{eq:14}
\rho_{AB|\Pi_{B}^{j}}= \frac{1}{p_{j}}
\left(\openone_{2}\otimes\Pi_{B}^{j}\right)\rho_{AB}\left(\openone_{2}\otimes\Pi_{B}^{j}\right),
\end{equation}
siendo $p_j$ la probabilidad de obtener dicho estado $\rho_{A|\Pi_{B}^{j}}$, esto es,
\begin{equation}\label{eq:15}
p_{j}=\tr{(\openone_{2}\otimes\Pi_{B}^{j})\rho_{AB}}.
\end{equation}
\noindent La entropía condicional despu\'{e}s de la medida ser\'{a}
$S\left(\rho_{AB}|\Pi_{B}^{j}\right)= \sum_{j}p_{j}S\left(\rho_{A}|j\right)$, siendo
$S\left(\rho_{A}|j\right) = \textrm{Tr}_{B}(\rho_{AB|\Pi_{B}^{j}})$ la entropía del subsistema $A$
despu\'{e}s de la medida. La informaci\'{o}n mutua del sistema $AB$ despu\'{e}s de la medida
ser\'{a}:
\begin{equation}\label{eq:16}
J\left(\rho_{AB}\right) = S\left(\rho_{A}\right)- S\left(\rho_{AB}|\Pi_{B}^{j}\right) \ .
\end{equation}
La medida de la discordia cu\'{a}ntica se define como
\begin{equation}\label{eq:17}
\delta_{\overleftarrow{AB}}\left(\rho_{AB}\right) = I\left(\rho_{AB}\right)- \max_{\Pi_{B}} \
J\left(\rho_{AB}\right) \ .
\end{equation}
La maximizaci\'{o}n se realiza sobre todas las medidas posibles $\left\{\Pi_{B}^{j}\right\}$. Esta
discordia es siempre po\-si\-ti\-va, asim\'{e}trica (ya que
$\delta_{\overleftarrow{AB}}\left(\rho_{AB}\right)\neq\delta_{\overrightarrow{AB}}\left(\rho_{AB}\right)$),
es cero si y solo si las medidas locales no perturban al sistema cu\'{a}ntico, y es invariante
ante transformaciones unitarias. El cálculo de la DC exige en muchos casos conocidos un proceso de
maximizaci\'{o}n numérica de la entropía condicional, y en muy pocos casos se conoce la solución
analítica. En el presente trabajo determinamos analíticamente la DC para la matriz de cuasi-Werner
$\rho_{W}({\alpha}_{D},f(n),p)$ dada por la ecuaci\'{o}n \eqref{eq:12}, encontrando que esta tiene
la forma
\begin{figure*}[t]
\begin{picture}(0,4)(-1.5,0)
\put(0,0){\includegraphics[height=4.5cm,width=4.5cm]{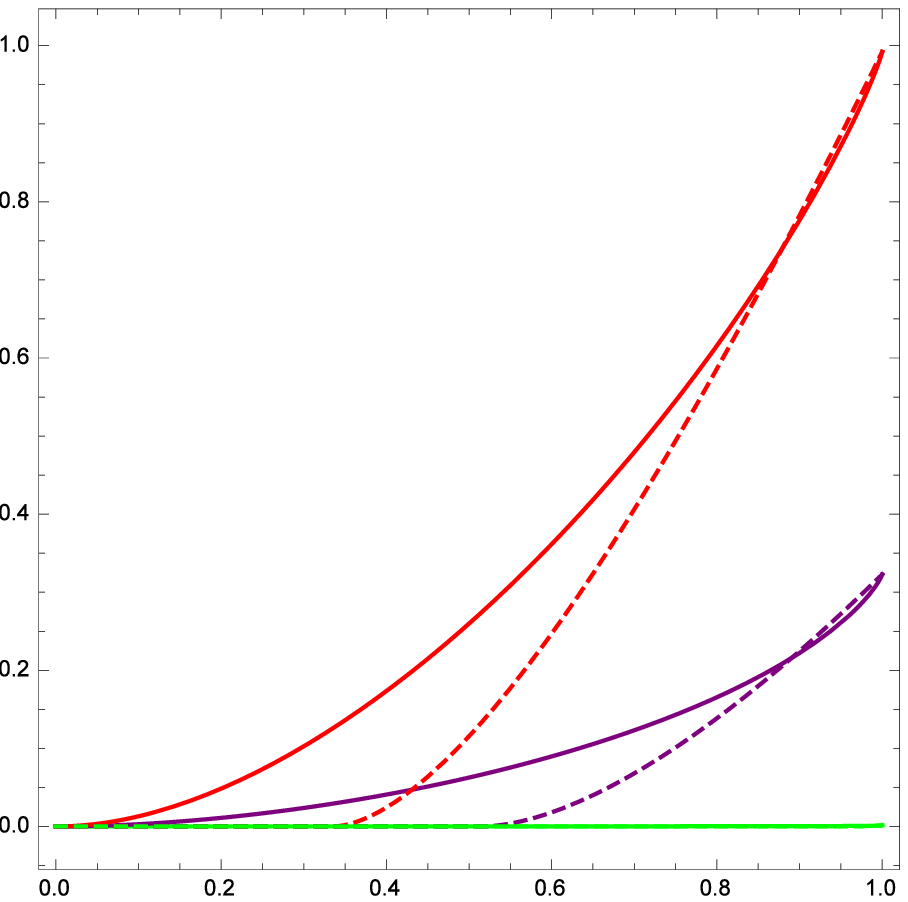}
           \put(-3,3){\footnotesize(a)} }
\put(5.5,0){\includegraphics[height=4.5cm,width=4.5cm]{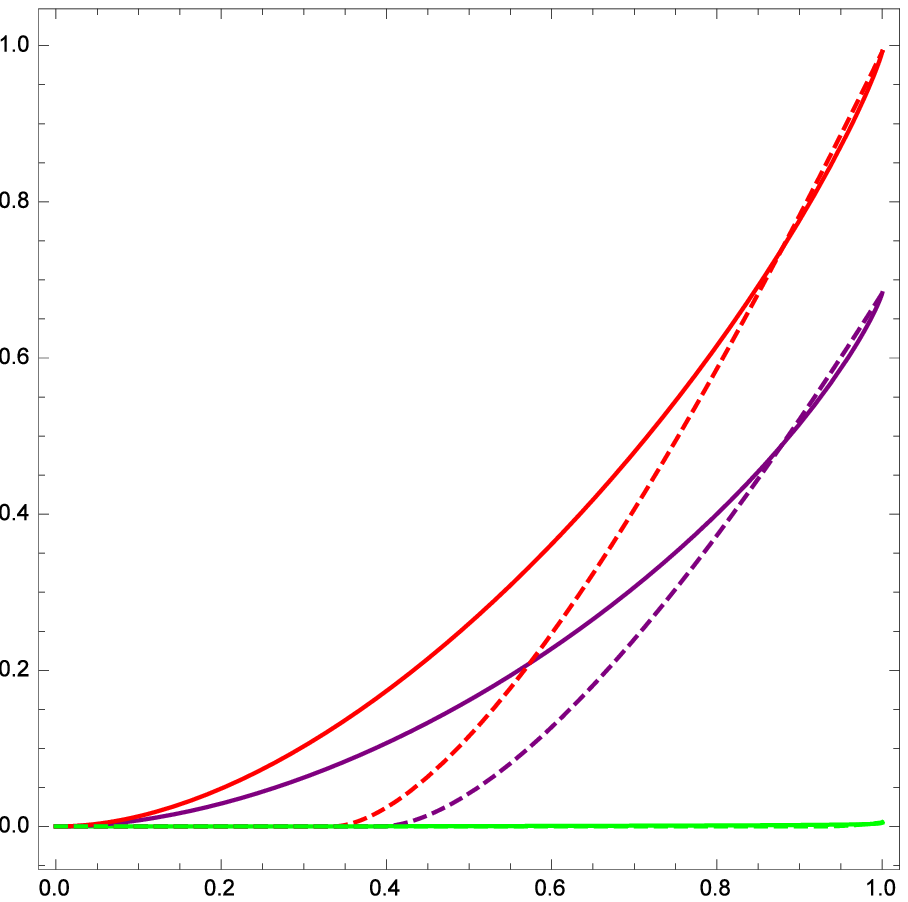}
          \put(-3,3){\footnotesize(b)} } 
\put(11,0){\includegraphics[height=4.5cm,width=4.5cm]{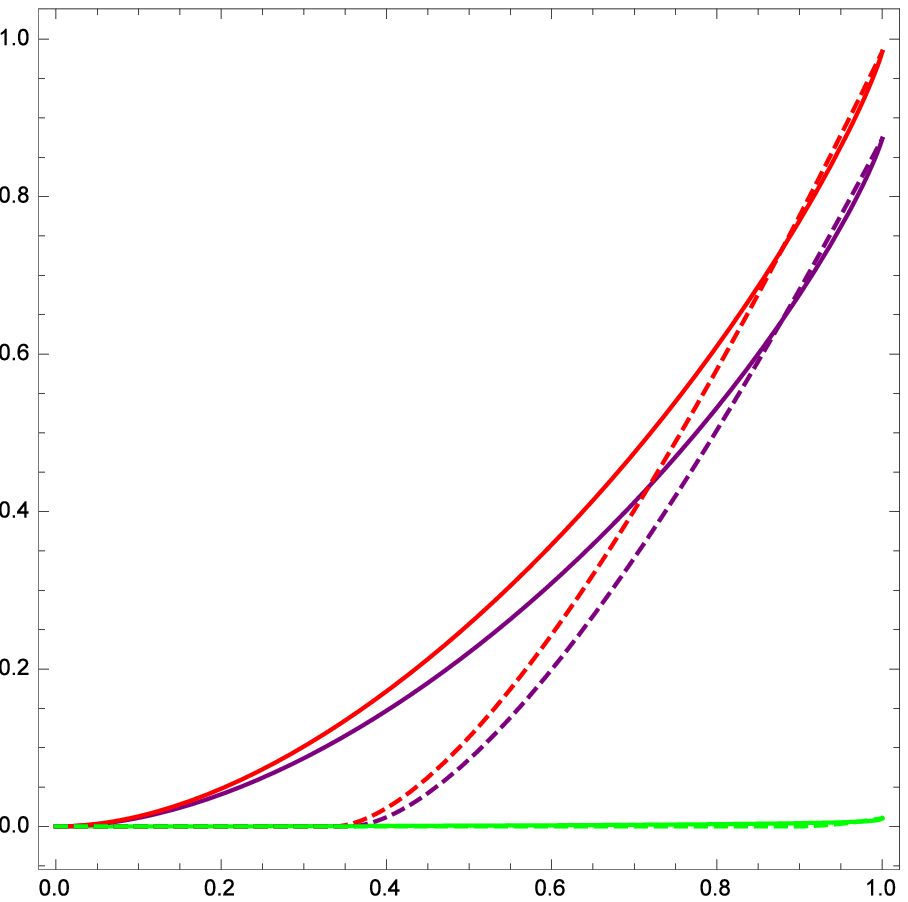}
            \put(-3,3){\footnotesize(c)} }
\end{picture}
\caption{Gráficas de la DC (linea solida) y EOF (linea segmentada) en función del parámetro de
mezcla $p$, con $\alpha=0.1$ (verde), $\alpha=0.5$ (morado) y $\alpha=1$ (rojo), para (a) $k=1/2$,
(b) $k=1$ y (c) $k=3/2$. } \label{fig:1}
\end{figure*}
\begin{equation}\label{eq:18}
\delta_{\overleftarrow{AB}}(\rho_{AB})=S(\rho_B)+F(z_+)+F(z_-)-S(\rho_{AB})
\end{equation}
con
\begin{subequations}
\label{eq:24}
\begin{align}
\label{eq:24a}%
z_\pm&=\frac{1}{2}\pm\frac{4q-1}{24}n_+^2\frac{N_-^4-N_+^4}{N_+^4N_-^4}
\\
\label{eq:24b}%
F(z)&=\tfrac{1-q}{3}\ln\left[\tfrac{3z}{1-q}-1\right]-z\ln\left[1-\tfrac{1-q}{3z}\right]
\end{align}
\end{subequations}
Mientras que la entropía $S(\rho_B)$ viene dada por
\begin{equation}\label{eq:25}
S(\rho_B)=h\left(\tfrac{1}{2}-\tfrac{1}{2}p\sqrt{1-C_\psi^2}\right)
\end{equation}
siendo $h(x)$ la función de entropía binaria y $C_\psi=n_+^2/2N_+^2N_-^2$ la concurrencia del
estado dado en \eqref{eq:11}.

\section*{Resultados y conclusiones}
En la Figura \ref{fig:1}, se muestra la relación entre la DC (líneas sólidas) y el EoF (líneas
segmentadas), en función del parámetro de mezcla $p$. En función de los resultados obtenidos hemos
definido, a partir de la amplitud $|\alpha|$, tres regiones de interés. La región con
$|\alpha|\leq 0.1$ (valores pequeños de $|\alpha|$), la región con $0.1 <|\alpha| < 1$ (valores
intermedios de $|\alpha|$),  y la región con $|\alpha|\geq 1$ (valores grandes de $|\alpha|$). En
cada una de las gráficas contenidas en la Figura \ref{fig:1} se comparan la DC y del EoF para los
valores de $|\alpha|$; $|\alpha|= 0.1$ (líneas verdes), $|\alpha|= 0.5$ (líneas moradas) y
$|\alpha|= 1$ (líneas rojas). La Figura \ref{fig:1}(a), corresponde al valor del parámetro de
deformación $k=1/2$, la Figura \ref{fig:1}(b), corresponde al valor del parámetro de deformación
$k=1$, y la Figura \ref{fig:1}(c), corresponde al valor del parámetro de deformación $k=3/2$.
\\
\par\noindent
La DC y EoF no son susceptibles a los efectos introducidos por la función de deformación para
va\-lo\-res pequeños y grandes del parámetro $|\alpha|$. Sin embargo, en la región de valores
intermedios de dicho parámetro (líneas moradas), se puede observar que la DC y EoF cambian
conforme lo hace el parámetro de deformación $k$. Además, a medida que $|\alpha|$ aumenta también
lo hace de forma monótona la DC y el EoF. Por otra parte, se puede observar que el EoF y la DC
coinciden cuando el parámetro de mezcla es $p=1$\cite{Ollivier2001,Henderson2001}, y alcanzan su
máximo valor. Esto se debe a que el estado \eqref{eq:12} es puro para dicho valor de $p$. También,
se puede notar que siempre existe DC aún en aquellas regiones de $p$ donde no existe
entrelazamiento (estados producto); por lo que la DC mide otro tipo de correlación mas allá del
entrelazamiento.
\\
\par\indent
En conclusión, se puede emplear la DC para los estados estudiados como recurso en protocolos de
computación cuántica y/o para el procesamiento cuántico de la información, ya que al modificar el
parámetro $k$, en la región de valores intermedios de $\alpha$, se puede calibrar en forma
discreta los valores de la DC.
\\
\par\indent
En el presente trabajo se obtuvo una expresión analítica de la DC y EoF para los estados tipo
Werner bipartitos entrelazados $f-$deformados, tomando como función de deformación la funciones de
Barut-Girardello y el estado tipo Bell $\psi^+$. Encontrando,

\begin{footnotesize}

\end{footnotesize}

\end{document}